\documentclass[12pt]{article}

\usepackage{amssymb}
\usepackage{subfig}
\usepackage{color}
\usepackage{epsfig,amssymb,amsfonts,amsmath,graphicx,dsfont,cite,xfrac}
\usepackage{authblk}
\definecolor{mygray}{gray}{0.5}

\usepackage{cite}
\usepackage[colorlinks=true,linkcolor=blue,citecolor=red]{hyperref}
\usepackage{tikz,pgfplots}
\usetikzlibrary{positioning}
\usetikzlibrary{matrix}

\begin{document}
\title{Nonstationary deformed singular oscillator: quantum invariants and the factorization method}

\author[${}$]{Kevin Zelaya\thanks{zelayame@crm.umontreal.ca}}

\affil[${}$]{\footnotesize Centre de Recherches Math\'ematiques, Universit\'e de Montr\'eal, Montr\'eal H3C 3J7, QC, Canada}

\date{}

%\ead{zelayame@crm.umontreal.ca}
\maketitle

\begin{abstract}
New families of time-dependent potentials related with the stationary singular oscillator are introduced. This is achieved after noticing that a non stationary quantum invariant can be constructed for the singular oscillator. Such invariant depends on coefficients that are related to solutions of an Ermakov equation, the latter becomes essential since it guarantees the regularity of the solutions at each time. In this form, after applying the factorization method to the quantum invariant, rather than the Hamiltonian, one manages to introduce the time parameter into the transformation, leading to factorized operators which are the constants of motion of the new time-dependent potentials. Under the appropriate limit, the initial quantum invariant reduces to the stationary singular oscillator Hamiltonian, in such case, one recovers the families of potentials obtained through the conventional factorization method and previously reported in the literature. In addition, some special limits are discussed such that the singular barrier of the potential vanishes, leading to non-singular time-dependent potentials.
\end{abstract}

\section{Introduction}
In quantum mechanics, the dynamic of non-relativistic systems is determined from the solutions of the Schr\"odinger equation. An essential part of the dynamical law is the Hamiltonian operator, which characterizes the system under consideration. In \textit{stationary cases} (time-independent), the Hamiltonian plays the role of the energy observable, and the initial mathematical problem is reduced to a simpler eigenvalue equation. Even in this case, only some few stationary Hamiltonians are known to admit exact solutions, such as the ones that describe the harmonic oscillator, hydrogen atom and the interaction between diatomic molecules, just to mention some. Thus, the search of new exactly solvable models becomes a mathematical challenge. In this regard, the factorization method~\cite{Mie04,Kha93,Car00,Coo01,Don07} becomes an outstanding technique to explore the existence and constrution of new exactly-solvable stationary models. The method is also explained from its relationship with the darboux transformation~\cite{Mat91}. In this form, a wide class of new exactly solvable models have been reported in the literature for Hermitian Hamiltonians~\cite{Mie84,Mie00,Fer99}, non-Hermitian Hamiltonians with real spectrum in the $\mathcal{PT}$ and non-$\mathcal{PT}$ regime~\cite{Zno00,Bag01,Ros15,Ros18,Bla18,Cor15}, position-dependent mass models~\cite{Que06,Cru09,Cru13}, among other cases. 

For \textit{nonstationary systems} (time-dependent), the Schr\"odinger equation, which is a partial differential equation, can not be reduced into an eigenvalue equation and must be solved directly. The latter brings some mathematical difficulties, and under some instances, one has to rely on approximation techniques such as the sudden and the adiabatic approximations~\cite{Sch02} to extract partial information from the system, if the required criteria are met. Despite its complexity, time-dependent phenomena find interesting applications in electromagnetic traps of charged particles~\cite{Pau90,Com86,Pri83,Gla92}, as well as in optical-analogs under the paraxial approximation~\cite{Cru17,Raz19,Con19}. Among the nonstationary quantum systems, the parametric oscillator~\cite{Dod95,Zel19} is perhaps the most well known model that admits a set of exact solutions. Lewis and Riesenfeld~\cite{Lew69} addressed the problem in the quantum case by noticing the existence of a nonstationary eigenvalue equation associated with the appropriate \textit{constant of motion} (\textit{quantum invariant}) of the system. 

In a similar form to the conventional factorization method, Bagrov and Samsonov~\cite{Bag95,Bag96} have proposed an alternative approach to construct new solvable time-dependent potentials. In the latter, a set of intertwining relationships that link two different Schr\"odinger equations is introduced. This relates a well known model with another one which is unknown and to be determined. The method has been proved to be useful in the construction of new families of exactly-solvable deformed nonstationary oscillators~\cite{Zel17,Con17,Cru19,Cru20}. However, the method by itself does not provide information about the constants of motion of the system, which have to be computed in a different way. Moreover, the solutions of the new model are not necessarily orthogonal, even if the initial model has orthogonal solutions, see~\cite{Zel17}.

In this work, the stationary singular oscillator~\cite{Gol61,Wei79,Cru11} is considered as the initial potential. This finds interesting application as a model that characterized two-ion traps~\cite{Dod98}. Therefore, it is desirable to find additional potentials with such a property. The latter is achieved if one departs from the nonstationary spectral problem related with the singular oscillator, which is determined from the appropriate quantum invariant of the system. Those nonstationary eigenfucntions satisfy a second order differential equation in the spatial variable, whose coefficients depend on time and are related to solutions of the Ermakov equation. From the latter, the time-dependence is inherited to the factorization method, leading to families of time-dependent potentials, even if the initial Hamiltonian is stationary.

The organization of this paper is as follows. In Sec.~\ref{sec:SO}, the solutions of the stationary singular oscillator are briefly discussed. Then, an additional quantum invariant, different from the Hamiltonian, is constructed and its spectral problem is properly identified. In Sec.~\ref{sec:TDSO}, the implementation of the factorization method on the aforementioned quantum invariant is introduced, leading to a new family of time-dependent potentials whose solutions are mapped from the initial system. Some particular cases are discussed in Sec.~\ref{sec:cases}, where it is shown how to recover, as a particular limit, the well known stationary results. Additional details on the construction of the new Hamiltonians are presented in App.~\ref{sec:APPA}. In App.~\ref{sec:APPB}, the intermediate steps required in the calculation of the normalization constant associated with the added eigenvalue are presented. Final comments and perspectives of this work are presented in Sec.~\ref{sec:conclu}.

\section{Singular Oscillator}
\label{sec:SO}
The stationary singular oscillator is defined through the Hamiltonian
\begin{equation}
\hat{H}_{1}=\hat{p}^{2}+V_{1}(\hat{x}) \, , \quad V_{1}(\hat{x})=\hat{x}^{2}+\frac{g(g+1)}{\hat{x}^{2}} \, ,
\label{eq:SO1}
\end{equation}
with $g\geq 0$ an arbitrary constant. Given that $\hat{H}_{1}$ is time-independent, the Schr\'odinger equation
\begin{equation}
i\frac{\partial}{\partial t}\psi(x,t)=\hat{H}_{1}\psi(x,t) \, ,
\label{eq:}
\end{equation}
admits a set of stationary orthonormal solutions $\{ \psi_{n}(x)\}_{n=0}^{\infty}$ computed through the time-evolution operator $\hat{U}(t)=e^{-i\hat{H}_{1}t}$ and the eigenvalue equation 
\begin{equation}
\hat{H}_{1}\phi_{n}(x)\equiv-\frac{\partial^{2}\phi_{n}(x)}{\partial x^{2}}+\left[x^{2}+\frac{g(g+1)}{x^{2}} \right]\phi_{n}(x)=E_{n}\phi_{n}(x) \, , \quad \psi_{n}(x,t)=e^{-iE_{n}t}\phi_{n}(x) \, .
\label{eq:SO3}
\end{equation}
where the coordinate representation $\hat{x}\rightarrow x$ and $\hat{p}\rightarrow -i\partial/\partial x$ has been used. The singular oscillator is one of the few exactly solvable models in quantum mechanics, and its eigenfunctions $\phi_{n}(x)$ and eigenvalues $E_{n}$ have been reported in the literature~\cite{Gol61,Cru11},
\begin{equation}
\phi_{n}(x)=\mathcal{N}_{n} e^{-\frac{x^{2}}{2}}x^{g+1}L_{n}^{(g+1/2)}\left(x^{2}\right) \, , \quad \mathcal{N}_{n}^{2}=2\frac{\Gamma(n+1)}{\Gamma(n+g+3/2)} \, , \quad E_{n}=4n+2g+3 \, .
\label{eq:eigenH1}
\end{equation}
with $L_{n}^{(m)}(z)$ the \textit{associated Laguerre polynomials}~\cite{Olv10}. The normalization constant $\mathcal{N}_{n}$ was fixed from the condition $\langle \phi_{n}\vert\phi_{n}\rangle=1$, where the physical inner-product is defined as
\begin{equation}
\langle f \vert g \rangle=\int_{0}^{\infty}dx f^{*}(x)g(x) \, ,
\label{eq:inn}
\end{equation}
with $z^{*}$ the complex conjugate of $z$ and $f(x)=\langle x\vert f\rangle$ is the coordinate representation of the vector $\vert f\rangle$.

\subsection{Nonstationary quantum invariant}
\label{subsec:quantI1}
Remarkably, even if the Hamiltonian is time-independent, there is a constant of motion, different from the Hamiltonian $\hat{H}_{1}$. The latter is a fact that was explored for the stationary oscillator~\cite{Zel19}, and it was used in the construction of new solvable time-dependent models~\cite{Cru20,Zel19b}. 

To illustrate the existence of such constant of motion, consider the time-dependent operator of the form
\begin{equation}
\hat{I}_{1}(t)=C_{0}(t)\left( \frac{\hat{p}^{2}}{2m}+\frac{g(g+1)}{\hat{y}^{2}}\right)+ C_{1}(t)\hat{y}^{2}+C_{2}(t)\{\hat{y},\hat{p} \} \, ,
\label{eq:INVI0}
\end{equation}
where $\{\hat{x},\hat{p} \}=\hat{x}\hat{p}+\hat{p}\hat{x}$ is the anti-commutation relationship and the real-valued functions $C_{i}(t)$, for $i=1,2,3$, are determined from the quantum invariant condition
\begin{equation}
\frac{d}{dt}\hat{I}_{1}(t)=i[\hat{H}_{1},\hat{I}_{1}(t)]+\frac{\partial}{\partial t}\hat{I}_{1}(t)=0 \, .
\label{eq:SO4}
\end{equation}
Notice that a particular solution should be given as $C_{0}=C_{1}=1$ and $C_{2}=0$, where the operator $\hat{I}_{1}(t)$ simply reduces to the Hamiltonian $\hat{H}_{1}$, which is indeed a constant of motion of the system. In the general case, with the use of the identities
\begin{equation}
\begin{aligned}
& \left[ \hat{y}^{2},\hat{p}^{2}+\frac{g(g+1)}{\hat{x}^{2}}\right]=2i\{\hat{x},\hat{p} \} \, , \quad 
[\hat{y}^{2},\hat{p}^{2}]=2i\hbar\{\hat{y},\hat{p} \} \, , \\
& \left[ \{\hat{x},\hat{p} \}, \hat{p}+\frac{g(g+1)}{\hat{y}^{2}}\right]=4i\hbar\left( \hat{p}^{2}+\frac{g(g+1)}{\hat{x}^{2}}\right) \, .
\end{aligned}
\label{eq:SO6}
\end{equation}
the following set of coupled equations are obtained:
\begin{equation}
2 (C_{1}-C_{0})+\dot{C}=0 \, , \quad 4 C_{2} + \dot{C}_{0}=0 \, , \quad -4C_{2}+\dot{C}_{1}=0 \, , \quad \dot{f}(t)=\frac{d f(t)}{d t} \, .
\end{equation}
The latter is solved with ease, but it is convenient to introduce the reparametrization $C_{0}(t)=\sigma^{2}(t)$ such that, after some calculations, one obtains
\begin{equation}
\ddot{\sigma}+4\sigma=\frac{1}{\sigma^{3}} \, , \quad C_{0}(t)=\sigma^{2} \, , \quad C_{1}(t)=\frac{\dot{\sigma}^{2}}{4}+\frac{1}{\sigma^{2}} \, , \quad C_{2}(t)=-\frac{\sigma\dot{\sigma}}{2} \, , 
\label{eq:erm}
\end{equation}
where $\sigma(t)$ solves the Ermakov equation~\cite{Erm08}. Such an equation admits a solution through the nonlinear combination\cite{Ros15,Bla18}
\begin{equation}
\sigma^{2}(t)=a q_{1}^{2}(t)+ b q_{1}(t)q_{2}(t)+c q_{2}^{2}(t) \, , \quad b^{2}-4ac=-\frac{16}{W_{0}^{2}} \, ,
\label{eq:sigma}
\end{equation}
with $W_{0}=q_{1}\dot{q}_{2}-\dot{q}_{1}q_{2}\not=0$ the Wronskian of two linearly independent solutions of the classical equation of motion $\ddot{q}_{1,2}+4q_{1,2}=0$. These two solutions are given by
\begin{equation}
q_{1}(t)=\cos[2(t-t_{0})] \, , \quad q_{2}(t)=\sin[2(t-t_{0})] \, , \quad W_{0}=2 \, , 
\label{eq:lineal}
\end{equation}
with $t_{0}$ an arbitrary real phase-shift. After some simplifications, the solution of the Ermakov equation takes the form
\begin{equation}
\sigma^{2}(t)=\frac{a+c}{2}+\frac{a-c}{2}\cos[4(t-t_{0})]+\sqrt{a c-1} \, \sin[4(t-t_{0})] \, ,
\label{eq:ermf}
\end{equation}
where the parameters $a,c>0$, together with the costarint in~\eqref{eq:sigma}, ensure that $\sigma(t)$ is a nodeless function for $t\in\mathbb{R}$, for details see~\cite{Bla18}. 

Now, from the Lewis-Riesenfeld approach~\cite{Lew69}, it follows that the quantum invariant $\hat{I}_{1}(t)$ solves an eigenvalue equation of the form
\begin{equation}
\hat{I}_{1}(t)\varphi^{(1)}_{n}(x,t)=\lambda^{(1)}_{n}\varphi^{(1)}_{n}(x,t) \, , 
\label{eq:SO13}
\end{equation}
where $\lambda^{(1)}_{n}$ are the time-independent eigenvalues~\cite{Lew69} and $\varphi_{n}^{(1)}(x,t)$ the \textit{nonstationary eiegnfunctions} which satisfy the finite-norm condition $\langle \varphi_{n}^{(1)}(t)\vert\varphi_{n}^{(1)}(t)\rangle<\infty$, with the inner product as defined in~\eqref{eq:inn}. It is worth to remark that $\varphi_{n}^{(1)}(x,t)$ are not solutions of the Schr\"odinger equation~\eqref{eq:SO1}, but they are used to construct the solutions $\psi_{n}^{(1)}(x,t)$ through the addition of the appropriate time-dependent complex-phase~\cite{Lew69}
\begin{equation}
\psi^{(1)}_{n}(x,t)=e^{i\theta^{(1)}_{n}(t)}\varphi^{(1)}_{n}(x,t) \, , \quad \frac{d}{dt}\theta^{(1)}_{n}(t)=\langle \varphi^{(1)}_{n}(t)\vert i\frac{\partial}{\partial t}-\hat{H}_{1}\vert\varphi_{n}^{(1)}(t)\rangle \, ,
\label{eq:SO13-1}
\end{equation}
where $\psi_{n}^{(1)}(x,t)$ are indeed solutions of the Schr\"odinger equation. Contrary to the stationary solutions $\psi_{n}(x)$ of~\eqref{eq:SO3}, the phase $\theta_{n}^{(1)}(t)$ is not related with the time evolution of the system, except for the cases in which $\hat{I}_{1}(t)=\hat{H}_{1}$. 

Before proceeding, it is required to solve the eigenvalue equation~\eqref{eq:SO13}. To this end, it is convenient to introduce the coordinate representation and the reparametrization
\begin{equation}
\varphi^{(1)}_{n}(x,t)=\frac{e^{i\frac{\dot{\sigma}}{4\sigma}x^{2}}}{\sqrt{\sigma}}\chi_{n}(z(x,t)) \, , \quad z=z(x,t)=\frac{x}{\sigma} \, .
\label{eq:SO14}
\end{equation}
Notice that the reparametrization $z(x,t)$ is well defined at each time, since it has been guaranteed that $\sigma(t)$ is a nodeless function at each time. After substituting~\eqref{eq:SO14} in~\eqref{eq:SO13}, one recovers a differential equation for $\chi_{n}(z(x,t))$ of the form 
\begin{equation}
-\frac{\partial^{2}\chi_{n}}{\partial z^{2}}+\left[ z^{2}+\frac{g(g+1)}{z^{2}} \right] \chi_{n}=\lambda^{(1)}_{n}\chi_{n} \, ,
\label{eq:chi1}
\end{equation}
where it is clear that $\chi_{n}(z)$ solves the same eigenvalue equation~\eqref{eq:SO3}, but in the $z$-parameter instead. One thus have
\begin{equation}
\chi_{n}(z)=\mathcal{N}_{n} e^{-z^{2}{2}}z^{g+1}L_{n}^{(g+1/2)}\left(z^{2}\right) \, , \quad \lambda^{(1)}_{n}=4n+2g+3  \, ,
\label{eq:SO15}
\end{equation}
with $\mathcal{N}_{n}$ the normalization constant given in~\eqref{eq:eigenH1}. Interestingly, the nonstationary eigenfunctions~\eqref{eq:SO14} have found applications in wave propagations optical models, where the wave-packets are described by self-focusing Laguerre-Gaussian modes~\cite{Cru20b}. 

\begin{figure}
\centering
\subfloat[][$n=0$]{\includegraphics[width=0.25\textwidth]{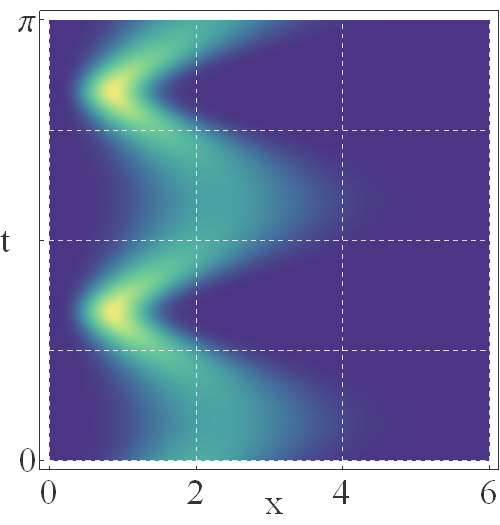}}
\hspace{5mm}
\subfloat[][$n=1$]{\includegraphics[width=0.25\textwidth]{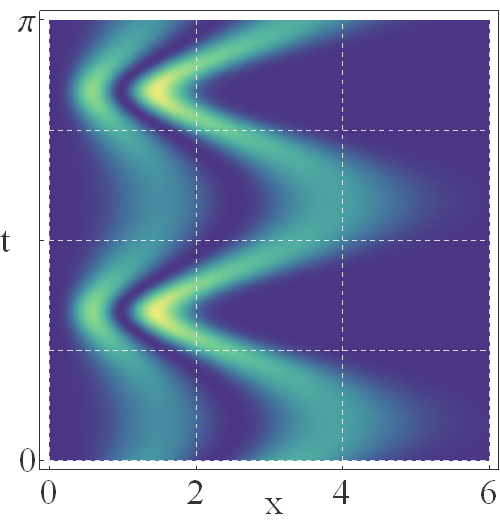}}
\hspace{5mm}
\subfloat[][$n=2	$]{\includegraphics[width=0.25\textwidth]{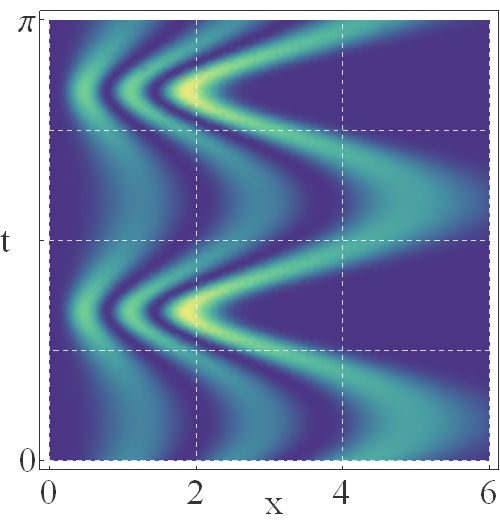}}
\caption{Nonstationary probability density $\vert \psi^{(1)}_{n}(x,t)\vert^2$=\eqref{eq:SO14} for $g=1$, $a=2$, $c=1$ and $t_{0}=0$.}
\label{fig:INV0}
\end{figure}

The reparametrization $z(x,t)$ also simplifies the calculations of the complex-phase $\theta^{(1)}_{n}(t)$ in~\eqref{eq:SO13-1}, leading to
\begin{equation}
\theta^{(1)}_{n}(t)=-\lambda^{(1)}_{n}\int^{t}\frac{dt'}{\sigma^{2}(t')}=-\frac{\lambda^{(1)}_{n}}{2}\arctan\left(\sqrt{ac-1}+c \tan[2(t-t_{0})] \right) \, ,
\label{eq:SO16}
\end{equation}
where the integral has been solved using the properties of the Ermakov equation, for details see~\cite{Bla18}. It is worth to mention that the orthogonality of the set $\{ \psi^{(1)}_{n}(x,t) \}_{n=0}^{\infty}$ holds provided that the solutions are evaluated at the same time, that is, $\langle\psi_{m}^{(1)}(t)\vert\psi_{n}^{(1)}(t)\rangle=\delta_{n,m}$. For different times though, the orthogonality does not longer hold, $\langle\psi_{m}^{(1)}(t')\vert\psi_{n}^{(1)}(t)\rangle\neq\delta_{n,m}$. From the completeness of the associated Laguerre polynomials, it is guaranteed that the nonstationary solutions $\psi_{n}^{(1)}(x,t)$ form a complete set of solutions, with a well defined number of zeros and interlacing properties at each time. The respective probability densities $\vert\psi^{(1)}_{n}(x,t)\vert^{2}$ are depicted in Fig.~\ref{fig:INV0} for $n=0,1,2$.
%----------- Figure: Probality density --------------

%-------------------------------------------------------

%%%%%%%%%%%%%%%%%%%%%%%%%%%%%%%%%%%%%%%%%%%%%%%%%%%%%%%%%%%%%%%%%%%%

%%%%%%%%%%%%%%%%%%%%%%%%%%%%%%%%%%%%%%%%%%%%%%%%%%%%%%%%%%%%%%%%%%%%
\section{Nonstationary deformed singular oscillator}
\label{sec:TDSO}
The time-dependent quantum invariant of the previous section, along with its respective set of nonstationary eigenfunctions, provides an alternative set of solutions to the Schr\"odinger equation of the singular oscillator. Those results can be used further in an attempt to construct new exactly solvable model. In previous works, the factorization method has been applied to the stationary singular oscillator to construct new families of stationary Hamiltonians, in such a way that the spectrum is preserved, with the exception of a possible added level~\cite{Fer13}. In this section, an alternative factorization is explored such that, even if the initial Hamiltonian is time-independent, the new resulting Hamiltonians are in general time-dependent. This is achieved by applying the factorization method to the quantum invariant $\hat{I}_{1}(t)$ instead of the Hamiltonian $\hat{H}_{1}$. The latter procedure has been proved useful while exploring the time-dependent rational extensions of the parametric oscillator~\cite{Zel19b}.

Consider a set of mutually adjoint operators defined in coordinate representation as first-order differential operators in the spatial variable~\cite{Zel19}, that is,
\begin{equation}
\hat{A}(t):=\sigma\frac{\partial}{\partial x}+w(x,t) \, , \quad \hat{A}^{\dagger}(t):=-\sigma\frac{\partial}{\partial x}+w^{*}(x,t) \, ,
\label{eq:int1}
\end{equation}
where $\sigma$ is solution of the Ermakov equation~\eqref{eq:ermf}. The complex-valued function $w(x,t)$ is determined from the factorization condition 
\begin{equation}
\hat{I}_{1}:=\hat{A}^{\dagger}(t)\hat{A}(t)+\epsilon \, ,
\label{eq:factA}
\end{equation}
where $\epsilon$ is a real constant and $\hat{I}_{1}(t)=$\eqref{eq:INVI0} is rewritten in coordinate representation as
\begin{equation}
\hat{I}_{1}(t)\equiv-\sigma^{2}\frac{\partial^{2}}{\partial x^{2}}+i\sigma\dot{\sigma}x\frac{\partial}{\partial x}+ R(x,t)+\sigma^{2}\frac{g(g+1)}{x^{2}} \, , \quad R(x):=i\frac{\sigma\dot{\sigma}}{2}+\left(\frac{\dot{\sigma}^{2}}{4}+\frac{1}{\sigma^{2}} \right)x^{2} \, .
\label{eq:invI1r}
\end{equation}
After substituting~\eqref{eq:int1} in~\eqref{eq:factA} and comparing with~\eqref{eq:invI1r} one obtains
\begin{equation}
w(x,t)=-i\frac{\dot{\sigma}}{2}x+W(x,t) \, , \quad -\sigma \frac{\partial W}{\partial x}+W^{2}=\frac{x^{2}}{\sigma^{2}}+\sigma^{2}\frac{g(g+1)}{x^{2}}-\epsilon \, ,
\label{eq:ricc1}
\end{equation}
where $W(x,t)$ is a real-valued function. Notice that the reparametrization $z=x/\sigma$ leads to a Riccati equation of the form
\begin{equation}
-\frac{\partial W}{\partial z}+W^{2}=z^{2}+\frac{g(g+1)}{z^{2}}-\epsilon \, ,
\label{eq:ricc2}
\end{equation}
where $W\equiv W(z(x,t))$ becomes a function of $z$ and it is solved though the linear equation
\begin{equation}
-\frac{\partial^{2}}{\partial z^{2}}u(z)+ \left[z^{2}+ \frac{g(g+1)}{z^{2}} \right]u(z)=\epsilon u(z) \, , \quad W(z)=-\frac{1}{u(z)}\frac{\partial u(z)}{\partial z} \, .
\label{eq:ricc3}
\end{equation}
The latter coincides with the spectral problem associated with the stationary singular oscillator. But in this case, the solutions $u(z)$ are not required to have a finite-norm. Nevertheless, given the relationship between $W(z)$ and $u(z)$, it is necessary to impose $u(z(x,t))$ to be a nodeless function in $x\in\mathbb{R}^{+}$, in such a way that $W(z)$ is a regular function in the same domain. In general, the solutions of~\eqref{eq:ricc3} are determined by taking~\eqref{eq:ricc3} into the hypergeometric differential equation form~\cite{Nik88}, leading to a general solution of the form
\begin{equation}
u(z)= \frac{e^{-\frac{z^{2}}{2}}}{z^{g}}\left[ k_{a}  \, z^{2g+1} \, {}_{1}F_{1}\left( \frac{3+2g-\epsilon}{4}, \frac{3}{2}+g;z^{2} \right)+ k_{b} \, {}_{1}F_{1}\left( \frac{1-2g-\epsilon}{4}, \frac{1}{2}-g ; z^{2}\right) \right] \, ,
\label{eq:solU}
\end{equation}
where ${}_{1}F_{1}(\cdot,\cdot;z)$ stands for the \textit{confluent hypergeometric function}~\cite{Olv10}. The arbitrary real constants $k_{a}$, $k_{b}$ and $\epsilon$ are constrained such that $u(z)$ satisfy the nodeless condition. A first condition is given by $\epsilon<\lambda_{0}^{(1)}$, this guarantees that the linear combination of the confluent hypergeometric functions in~\eqref{eq:solU} have at most one zero in $x\in\mathbb{R}^{+}$. Then, with the use of the asymptotic behavior of the confluent hypergeometric function~\cite{Olv10}, a relationship between $k_{a}$ and $k_{b}$ is determined such that the aforementioned zero is placed at $x\rightarrow\infty$. After some calculations one determined the conditions
\begin{equation}
\frac{k_{a}}{k_{b}} > -\frac{\Gamma\left(\frac{1}{2}-g \right)\Gamma\left(\frac{3+2g-\epsilon}{4} \right)}{\Gamma\left(\frac{3}{2}+g \right)\Gamma\left(\frac{1-2g-\epsilon}{4} \right)} \, , \quad \epsilon < 2g+3 \, \quad k_{b}\neq 0\, , .
\label{eq:constraint}
\end{equation}
With~\eqref{eq:constraint} the solutions to the Riccati equation $W(z(x,t))$ are free of singularities, except perhaps in $x\rightarrow 0$. 

Now, with the factorization operators $\hat{A}(t)$ and $\hat{A}^{\dagger}(t)$ already determined, it is convenient to introduce a new operator that is factorized as
\begin{equation}
\hat{I}_{2}(t):=\hat{A}(t)\hat{A}^{\dagger}(t)+\epsilon \, ,
\label{eq:invI2}
\end{equation}
which in coordinate representation takes the form
\begin{equation}
\hat{I}_{2}\equiv-\sigma^{2}\frac{\partial^{2}}{\partial x^{2}}+i\dot{\sigma}\sigma x \frac{\partial}{\partial x}+R(x)+\frac{g(g+1)}{z(x,t)^{2}}+F(z(x,t)) \, ,
\label{eq:invI2x}
\end{equation}
where 
\begin{equation}
F(z(x,t))=2\frac{\partial}{\partial z}W(z(x,t))=-2\frac{\partial^{2}}{\partial z^{2}}\ln u(z(x,t)) \, , .
\label{eq:pseusoV2}
\end{equation}
It is clear that $\hat{I}_{2}(t)$ is not a quantum invariant of the singular oscillator. However, one may determine the respective Hamiltonian $\hat{H}_{2}(t)$ for which $\hat{I}_{2}(t)$ is its quantum invariant. Such Hamiltonian is given as (see App.~\ref{sec:APPA} and~\cite{Zel19b} for details)
\begin{equation}
\hat{H}_{2}(t)\equiv -\frac{\partial^{2}}{\partial x^{2}}+V_{2}(x,t) \, , \quad V_{2}(x,t)=x^{2}+\frac{g(g+1)}{x^{2}}+\frac{1}{\sigma^{2}(t)}F(z(x,t)) \, .
\label{eq:potV2}
\end{equation}
Notice that, in general, the time-dependent potential $V_{2}(x,t)$ is not trivially separable as the sum of a spatial part plus a time-dependent part. Thus, the solutions of the Schr\"odinger equation may not be determined in a straightforward way if one tries to solve it directly. Nevertheless, in the sequel it is shown that the factorization operators lead to a mechanism to compute the solutions through simple mappings.

\subsection{Spectral properties of $\hat{I}_{2}(t)$ and solutions of $\hat{H}_{2}(t)$}
\label{subsec:specprop}
Now, with the new time-dependent Hamiltonian $\hat{H}_{2}(t)$ already identified, the solutions of the respective Schr\"odinger equation must be addressed. As discussed in Sec.~\ref{subsec:quantI1}, it is required to solve the spectral problem associated with $\hat{I}_{2}(t)$, and then the appropriate time-dependent complex-phase must be added to the nonstationary eigenfunctions. Remarkably, the spectral problem 
\begin{equation}
\hat{I}_{2}(t)\varphi_{n}^{(2)}(x,t)=\lambda_{n}^{(2)}\varphi_{n}^{(2)}(x,t) \, ,
\label{eq:mapI2}
\end{equation}
with $\varphi^{(2)}_{n}(x,t)$ and the $\lambda_{n}^{(2)}$ the respective nonstationary eigenfunctions and eigenvalues, is determined from the \textit{intertwining relationships} between $\hat{I}_{1}(t)$ and $\hat{I}_{2}(t)$. The latter is obtained from the factorizations defined in~\eqref{eq:factA} and~\eqref{eq:invI2}, leading to
\begin{equation}
\hat{I}_{1}(t)\hat{A}^{\dagger}(t)=\hat{A}^{\dagger}(t)\hat{I}_{2}(t) \, , \quad \hat{I}_{2}(t)\hat{A}(t)=\hat{A}(t)\hat{I}_{1}(t) \, .\label{eq:intert}
\end{equation}
Eq.~\eqref{eq:intert} provides a mechanism to map the eigenfunctions of $\hat{I}_{1}(t)$ into eigenfunctions of $\hat{I}_{2}(t)$, and vice versa. Also, it also allows to determine the respective eigenvalues $\lambda_{n}^{(2)}$ in terms of $\lambda_{n}^{(1)}$. In Sec.~\ref{subsec:quantI1}, the spectral problem related to $\hat{I}_{1}(t)$ was already identified. Thus, it is straightforward to obtain the spectral information for $\hat{I}_{2}(t)$ as
\begin{equation}
\varphi_{n+1}^{(2)}(x,t)=\frac{1}{\sqrt{\lambda_{n}^{(1)}-\epsilon}}\hat{A}(t)\varphi_{n}^{(1)}(x,t) \, , \quad \lambda_{n+1}^{(2)}=\lambda_{n}^{(1)} \, , \quad n=0,1\cdots \, ,
\label{eq:specI2-1}
\end{equation} 
where the orthogonality condition $\langle\varphi_{m}^{(2)}(t)\vert\varphi^{(2)}_{n}(t)\rangle=\delta_{n,m}$ is inherited from that of the set $\{ \varphi_{n}^{(1)}(x,t) \}$, with respect to the physical inner-product~\eqref{eq:inn}. The additional factor $(\lambda_{n}^{(1)}-\epsilon)^{-1/2}$ has been introduced as a normalization constant. Notice that the index of the mapped eigenfunctions in~\eqref{eq:specI2-1} has been fixed at $n+1$, this is because of the existence of an additional nonstationary eigenfunction $\varphi_{\epsilon}^{(2)}(x,t)$ that can not be constructed through the mapping provided by $\hat{A}(t)$. Such an eigenfunction, usually called \textit{missing state}, is well known in the literature about the factorization method for stationary systems~\cite{Kha93,Mie84,Mie04}. The missing state is determined from the orthogonality condition $\langle \varphi_{n+1}^{(2)}(t)\vert \varphi_{\epsilon}^{(0)}(t)\rangle=0$ for all $n=0,1,\cdots$. In this form it is guaranteed that $\varphi_{\epsilon}^{(2)}(x,t)$ is not a linear combination of the eigenfunctions in~\eqref{eq:specI2-1}, and thus it should be added to the set of elementary solutions, provided that $\varphi_{\epsilon}^{(2)}(x,t)$ satisfy the finite-norm condition. Straightforward calculations show that the orthogonality condition implies $\hat{A}^{\dagger}(t)\varphi_{\epsilon}^{(2)}(x,t)=0$, which also means that $\hat{I}_{2}(t)\varphi_{\epsilon}^{(2)}=\epsilon\varphi_{\epsilon}^{(2)}(x,t)$. That is, $\epsilon$ is an eigenvalue of the new quantum invariant and, from the nodeless condition $\epsilon < 2g+3 = \lambda_{0}^{(1)}$, it is the lowest eigenvalue in the spectrum. Thus, after some calculations, the normalized nonstationary eigenfunction $\varphi_{\epsilon}^{(2)}(x,t)$ takes the form 
\begin{equation}
\varphi_{0}^{(2)}(x,t)=\varphi_{\epsilon}^{(2)}(x,t)=\frac{\mathcal{N}_{\epsilon}}{\sqrt{\sigma}} \frac{e^{-i\frac{\dot{\sigma}}{4\sigma}x}}{u(z(x,t))} \, , \quad \lambda_{0}^{(2)}=\epsilon \, ,
\label{eq:missing}
\end{equation} 
where $\mathcal{N}_{\epsilon}$ stands for the normalization constant,  given as (see~\cite{Zel19T} and App.~\ref{sec:APPB} for details)
\begin{equation}
\mathcal{N}^{2}_{\epsilon}=(1+2g) \left[ k_{a}k_{b}+k^{2}_{b} \frac{\Gamma\left(\frac{1-2g}{2} \right)\Gamma\left(\frac{3+2g-\epsilon}{4}\right)}{\Gamma\left(\frac{3+2g}{2}\right)\left(\frac{1-2g-\epsilon}{4}\right)} \right] \, .
\label{eq:Nep}
\end{equation}
Eq.~\eqref{eq:Nep} holds provided that the constraints~\eqref{eq:constraint} are fulfilled. 

Now, following the discussion at the end of Sec.~\ref{sec:SO}, the nonstationary eigenfunctions of the quantum invariant are mapped into solutions of the Schr\"odinger equation $\psi_{n}^{(2)}(x,t)$ through the addition of the complex-phase
\begin{equation}
\psi_{n}^{(2)}(x,t)=e^{i\theta^{(2)}_{n}(t)}\varphi_{n}^{(2)}(x,t) \, , \quad \frac{d}{dt}\theta^{(2)}_{n}(t)=\langle \varphi_{n}^{(2)}(t)\vert i\frac{\partial}{\partial t}-\hat{H}_{2}(t)\vert\varphi_{n}^{(2)}(t)\rangle \, ,
\label{eq:schroI2}
\end{equation}
where $n=0,1,\cdots$. With the use of the reparametrization $z(x,t)=z/\sigma$, and after some calculations, one obtains (for details, see Appenddix A in~\cite{Zel19b})
\begin{equation}
\theta_{n}^{(2)}(t)=-\frac{\lambda_{n}^{(2)}}{2}\arctan\left( \sqrt{ac-1}+c\tan [2(t-t_{0})] \right) \, .
\end{equation}

\begin{figure}
\centering
\begin{tikzpicture}
  \matrix (m) [matrix of math nodes,row sep=3em,column sep=4em,minimum width=2em]
  {
     {i\frac{\partial}{\partial t}\psi^{(1)}=\hat{H}_{1}\psi^{(1)}}
     & {i\frac{\partial}{\partial t}\psi^{(2)}=\hat{H}_{2}(t)\psi^{(2)}} \\
     {\hat{I}_{1}(t)\varphi_{n}^{(1)}=\lambda_{n}^{(1)}\varphi_{n}^{(1)}} 
     & {\hat{I}_{2}(t)\varphi_{n}^{(2)}=\lambda_{n}^{(2)}\varphi_{n}^{(2)}} \\
     {\hat{I}_{1}(t)=\hat{A}^{\dagger}\hat{A}+\epsilon}
     & {\hat{I}_{2}(t)=\hat{A}\hat{A}^{\dagger}+\epsilon} \\ 
  };
  \path[-stealth]
    (m-1-1) edge [double] node [left] {$\psi_{n}^{(1)}=e^{i\theta^{(1)}_{n}(t)}\varphi^{(1)}_{n}$} (m-2-1)
    (m-2-1) edge [double] node [left] {} (m-3-1)
    (m-3-1) edge [double] node [below] {} (m-3-2)
    (m-3-2) edge [double] node [right] {} (m-2-2)
    (m-2-2) edge [double] node [right] {$\psi_{n}^{(2)}=e^{i\theta^{(2)}_{n}(t)}\varphi^{(2)}_{n}$} (m-1-2);
\end{tikzpicture}
\caption{Scheme summarizing the construction of $H_{2}(t)$, together with the respective solutions of the Schr\"odinger equation $\psi_{n}^{(2)}(x,t)$.}
\label{fig:F0}
\end{figure}

Therefore, both the spectral information of the invariant $\hat{I}_{2}(t)$ and the solutions of the Schr\"odinger equation associated with $\hat{H}_{2}(t)$ have been completely determined from the information of the initial stationary singular oscillator and its quantum invariant. It is worth to remark that, contrary to the stationary case, the factorization on the quantum invariant adds an additional level which is not necessarily an energy eigenvalue, nevertheless it provides a physical solution that can not be disregarded. A summary of the factorization method implemented in this work is depicted in the scheme of Fig.~\ref{fig:F0}.

%%%%%%%%%%%%%%%%%%%%%%%%%%%%%%%%%%%%%%%%%%%%%%%%%%%%%%%%%%%%%%%%%%%%

%%%%%%%%%%%%%%%%%%%%%%%%%%%%%%%%%%%%%%%%%%%%%%%%%%%%%%%%%%%%%%%%%%%%
\section{Particular cases}
\label{sec:cases}
\subsection{Stationary limit}
The condition $a=c=1$ leads to $\sigma(t)=1$ and consequently to $z=x$. In this limit, it is clear that $\hat{I}_{1}(t)\rightarrow\hat{H}_{1}$. Moreover, the nonstationary eigenfunctions converge to the eigenfunctions of the singular oscillator $\phi_{n}(x)$. The eigenvalues of both the Hamiltonian and the quantum invariant are the same, regardless of the stationary limit. In the same limit, both the new quantum invariant $\hat{I}_{2}(t)$ and the Hamiltonian $\hat{H}_{2}(t)$ converge the stationary models obtained through the conventional factorization method, already reported in the literature~\cite{Fer13}.

\begin{figure}
\centering
\subfloat[][]{\includegraphics[width=0.35\textwidth]{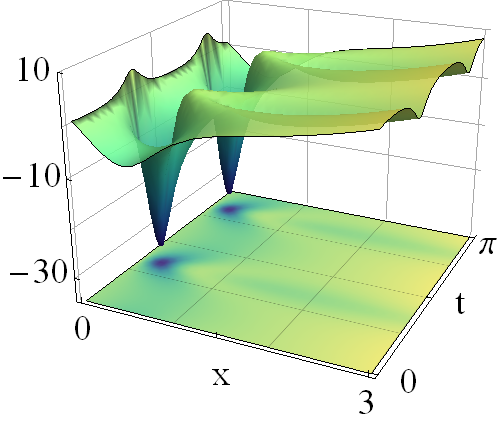}}
\caption{Non-singular potential $V_{2}(x,t)$ for $g=1$, $\epsilon=-2$, $k_{a}=1$, $k_{b}=1/4$, $a=2$, $c=1$ and $t=0$.}
\label{fig:V2g1}
\end{figure}

\subsection{Non-singular potentials $V_{2}(x,t)$}
It is worth to discuss the special class of potentials in $V_{2}(x,t)$ for which the singularity at $x=0$ is removed. A First case is obtained in the limit $g\rightarrow 0 $, the initial potential $V_{1}(x)$ reduces to the \textit{truncated oscillator}~\cite{Fer19}, which is a shape invariant case of the singular oscillator $V_{1}(x)=$~\eqref{eq:SO1}. In such a limit, the new potential $V_{2}(x,t)\vert_{g\rightarrow 0}$ is still time-dependent and non-singular at the origin. Clearly, in the stationary limit $a=c=1$, the potential $V_{2}(x,t)\vert_{g\rightarrow 0}$ reduces to the first-step transformed potentials reported in~\cite{Fer19}. 

A second way to remove the singularity at $x=0$ is achieved with $g=1$. In this case, the factorization method adds an additional singular term in the potential such that the singular-barrier vanishes. The behavior of $V_{2}(x,t)$ and the respective probability densities are depicted in Fig.~\ref{fig:V2g1}. In the latter, it can be seen that indeed the potential is finite at $x=0$ and its value changes periodically on time.

From the explicit form of the potential $V_{2}(x,t)$ it is easy to shoe that only the cases $g=0,1$ lead to non-singular potentials. It is worth to remark that, although the singularity has been removed, the domain of definition $x\in\mathbb{R}^{+}$ is still preserved. 

\subsection{Equidistant spectrum in $\hat{I}_{2}(t)$}
From~\eqref{eq:SO15}, it is clear that the spectrum of the initial quantum invariant $\hat{I}_{1}(t)$ is equidistant, $\lambda_{n+1}^{(1)}-\lambda_{n}^{(1)}=4$, for $n=0,1,\cdots$. The new quantum invariant $\hat{I}_{2}(t)$ admits equidistant spectrum if $\epsilon=2g-1$, which is physically admissible since it satisfies the constraint imposed in~\eqref{eq:constraint}. The respective eigenvalues of $\hat{I}_{2}(t)$ are then $\lambda_{n}^{(2)}=4n+2g-1$.

\begin{figure}
\centering
\subfloat[][$V_{2}(x,t)$]{\includegraphics[width=0.35\textwidth]{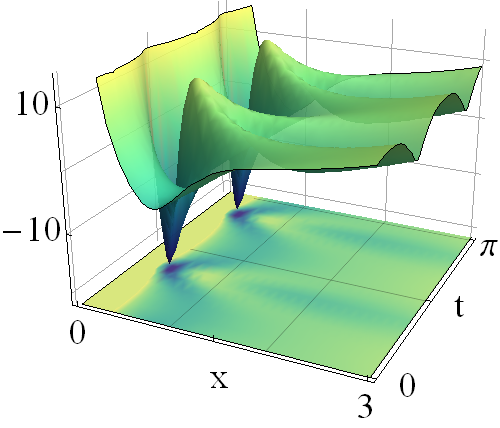}
\label{fig:Vwg2}}
\hspace{5mm}
\subfloat[][$V_{2}(x,t)$]{\includegraphics[width=0.35\textwidth]{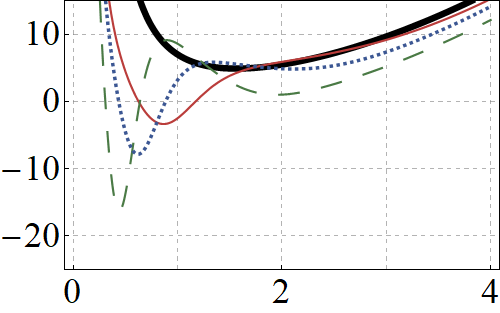}
\label{fig:Vwg21}}
\\
\subfloat[][$n=0$]{\includegraphics[width=0.25\textwidth]{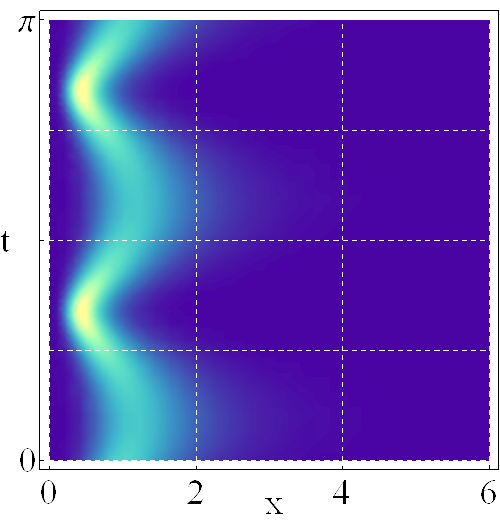}}
\hspace{5mm}
\subfloat[][$n=1$]{\includegraphics[width=0.25\textwidth]{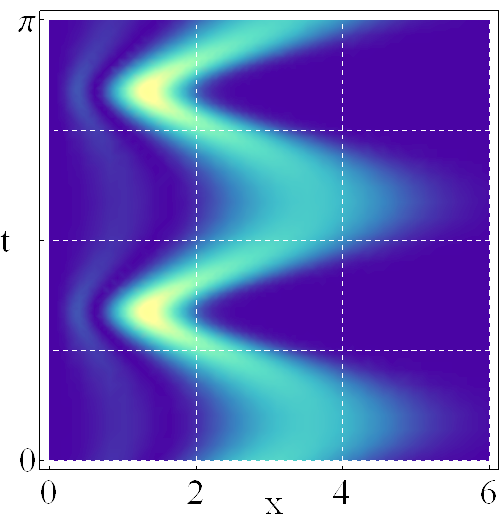}}
\hspace{5mm}
\subfloat[][$n=2$]{\includegraphics[width=0.25\textwidth]{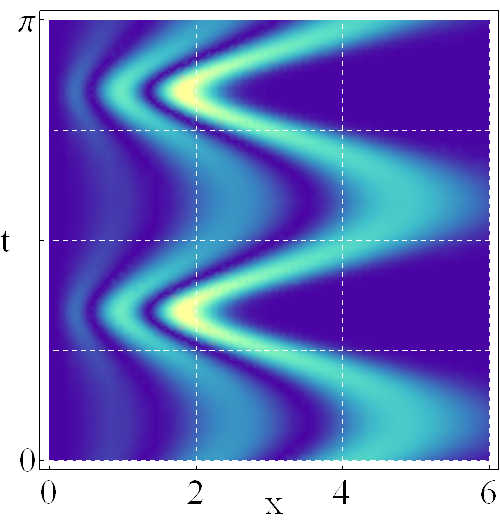}}
\caption{(a) Time-dependent potential $V_{2}(x,t)$ for $t\in[0,2\pi]$. (b) The 2-D projection of $V_{2}(x,t)$ for $t=0$ (solid-red), $t=3\pi/8$ (dashed-green) and $t=3\pi/3$ (dotted-blue) together with the stationary singular oscillator $V_{1}(x)$ (thick-black).
(Second row) Probability density $\vert\psi^{(2)}_{n}(x,t)\vert^{2}$ for the mentioned values of $n$. In all the cases the parameters are fixed as $\epsilon=3$, $g=2$, $a=1$, $c=2$, $t_{0}=0$, $k_{a}=1$ and $k_{b}=1/4$.}
\label{fig:WF2}
\end{figure}

To illustrate the form of the new potentials, the parameter are fixed at $g=2$ and $\epsilon=3$. Thus, one has
\begin{equation}
V_{2}(x,t)=x^{2}+\frac{2}{x^{2}}-\frac{2}{\sigma^{2}}\left(1+\frac{\partial^{2}}{\partial z^{2}}\ln \left[15\sqrt{\pi}k_{a}\textnormal{Erf}(z) + 8k_{b}-10k_{a}z(3+2z^2)e^{-z^2}\right] \right) \, ,
\label{eq:equiU}
\end{equation}
where $z=x/\sigma$ and the constants are constrained to $k_{b}>-\Gamma(7/2)k_{a}$. That is, the new potential and its solutions are well defined for any positive constants $k_{a}$ and $k_{b}$. The behavior of the potential and the respective probability densities are depicted in Fig.~\ref{fig:WF2}. From Fig.~\ref{fig:Vwg21}, the new time-dependent potential can be compared to the initial singular oscillator. Its clear that the minimum of the potential $V_{2}(x,t)$ is always lower to that of $V_{1}(x)$, as expected since the new time-dependent potential admits a new eigenvalue at $\epsilon<\lambda_{0}^{(1)}$. Asymptotically, at $x\rightarrow\infty$, both potentials seem to converge to the same, as the deformation produced by the factorization method is localized around the origin for the parameters used in this particular case. 

%%%%%%%%%%%%%%%%%%%%%%%%%%%%%%%%%%%%%%%%%%%%%%%%%%%%%%%%%%%%%%%%%%%%

%%%%%%%%%%%%%%%%%%%%%%%%%%%%%%%%%%%%%%%%%%%%%%%%%%%%%%%%%%%%%%%%%%%%
\section{Conclusions}
\label{sec:conclu}
It has been show that the proper implementation of the factorization method can be used to construct new time-dependent potentials. In contradistinction to the stationary cases, the factorization method is applied to the eigenvalue problem related with the appropriate time-dependent quantum invariant of singular oscillator. Even though the initial model is stationary, the departing point is the nonstationary quantum invariant, in this form the time-dependence is inherited to the new systems and its solutions through the solutions of a Ermakov equation. The factorization method leads to a natural way to obtain the respective constant of motion of the new system, whose spectral problem is also inherited from the former invariant. In this form, the existence of an orthogonal set of solutions for $\hat{H}_{2}(t)$ is justified, even if $\hat{H}_{2}(t)$ does not admit an spectral problem. It is worth to remark that the spectral analysis is carried out on the quantum invariant and not on the Hamiltonians. Nevertheless, the relationship between the nonstationary eigenfunctions and the solutions of the Schr\"odinger equation is simply given by a time-dependent complex-phase. 

In general, the potentials obtained so far are periodic functions on time. The latter is a consequence of the solution of the Ermakov equation. Such a class of potentials find interesting applications in the development of traps of particles, where the it is required a periodic potential on time to ensure that particles will be constrained in a certain region of the space~\cite{Gla92}. Additionally, the generalization of the results presented in this text are evident once the stationary limit is performed, where potentials previously reported in the literature are recovered~\cite{Fer13}. 

The construction presented in this text is quite general, but there are some special cases that deserve special attention on their own. For instance, the construction of the rational extensions of the stationary singular oscillator has been studied in the literature~\cite{Que11,Mar13}, where the set of solution is given in terms of exceptional Laguerre polynomials. Therefore, it is natural to explore such a construction under the time-dependent regime. The latter is something already achieved for the parametric oscillator~\cite{Zel19b}. In addition, the ladder operator structure can be addressed by exploiting the structure of the nonstationary Laguerre polynomials reported in Sec.~\ref{sec:SO}. In this form, the construction of coherent states could be determined with ease. These results are in preparation and will be reported elsewhere.

%%%%%%%%%%%%%%%%%%%%%%%%%%%%%%%%%%%%%%%%%%%%%%%%%%%%%%%%%%%%%%%%%%%%

%%%%%%%%%%%%%%%%%%%%%%%%%%%%%%%%%%%%%%%%%%%%%%%%%%%%%%%%%%%%%%%%%%%%
\appendix
\section{Determining $\hat{H}_{2}(t)$}
\label{sec:APPA}
The new time-dependent Hamiltonian $\hat{H}_{2}(t)$ related to the quantum invariant $\hat{I}_{2}(t)$ is determined from the following ansatz:
\begin{equation}
\hat{H}_{2}(t):=\hat{H}_{1}+G(t)F(z(\hat{x},t)) \, , 
\label{eq:APPA1}
\end{equation}
where $F(z(\hat{x},t))$ is defined in~\eqref{eq:pseusoV2} and $G(t)$ is determined from the quantum invariant condition
\begin{equation}
\frac{d}{dt}\hat{I}_{2}(t)=i[\hat{H}_{2}(t),\hat{I}_{2}(t)]+\frac{\partial}{\partial t}\hat{I}_{2}(t)=0 \, .
\end{equation}
With aid of the commutation relationships~\eqref{eq:SO6}, together with the identity $[\{\hat{x},\hat{p}\},T(\hat{x},t)]=2x[\hat{p},T(\hat{x},t)]$, valid for any smooth function $T(x,t)$ in the real $x$-variable, one obtains 
\begin{equation}
-i\left(G-\frac{1}{\sigma^{2}} \right)\left\{ \sigma^{2}[\hat{p}^{2},F(z(\hat{x},t))]-\sigma\dot{\sigma}\hat{x}[\hat{p},F(z(\hat{x},t))] \right\}=0 \, .
\label{eq:APPA2}
\end{equation}
It is clear that $G(t)=\sigma^{-2}(t)$ uniquely solves~\eqref{eq:APPA2}. With $G(t)$ and~\eqref{eq:APPA1}, the time-dependent Hamiltomian $\hat{H}_{2}(t)$ is then given as in~\eqref{eq:potV2}.

\section{Computing $\mathcal{N}_{\epsilon}$}
\label{sec:APPB}
The normalization constant $\mathcal{N}_{\epsilon}$ in~\eqref{eq:Nep} is determined from the finite-norm condition and the reparametrization $z(x,t)=x/\sigma$, leading to the integral
\begin{equation}
1=\vert\mathcal{N}_{\epsilon}\vert^{2}\int_{0}^{\infty}\frac{dz}{\left( k_{a}u_{1}(z)+k_{b} u_{2}(z) \right)^2}=\vert\mathcal{N}_{\epsilon}\vert^{2}\int_{0}^{\infty}\frac{dz}{u_2^2(z)}\frac{1}{\left[ k_{a} u_{1}(z)/u_{2}(z) + k_{b} \right]^2} \, ,
\label{eq:HNC4}
\end{equation}
where
\begin{equation}
u_{1}(z)= e^{-\frac{z^{2}}{2}}z^{g+1} \, {}_{1}F_{1}\left( \frac{3+2g-\epsilon}{4}, \frac{3}{2}+g;z^{2} \right) \, , \, u_{2}(z)= e^{-\frac{z^{2}}{2}}z^{-g}\, {}_{1}F_{1}\left( \frac{1-2g-\epsilon}{4}, \frac{1}{2}-g ; z^{2}\right) \, ,
\end{equation}
are the two linearly independent solutions of~\eqref{eq:ricc3}, with $\tilde{W}=u_{1}(z)\partial_{z}u_{2}(z)-u_{2}(z)\partial_{z}u_{1}(z)$ the respective Wronskian. Given that~\eqref{eq:ricc3} is an incomplete second order differential equation, it is strightforward to realize that $\tilde{W}$ is in general a constant~\cite{Zel19T}. Now, the change of variable $w=u_{1}/u_{2}$, together with $dz=-\tilde{W}/u_{2}^2(z)$, leads to
\begin{equation}
\frac{1}{\vert\mathcal{N}_{\epsilon}\vert^2}=-\frac{1}{\tilde{W}}\int_{w(z=0)}^{w(z\rightarrow\infty)}\frac{dw}{(k_{a}w+k_{b})^2}=\frac{1}{\tilde{W}k_{a}}\left[ \left.\frac{1}{k_{a}u_1(z)/u_{2}(z)+k_{b}}\right\vert_{0}^{\infty} \right] \, .
\label{eq:HNC5}
\end{equation}
From~\eqref{eq:HNC5} and the asymptotic behavior of the confluent hypergeometric function~\cite{Olv10} one recovers the normalization constant in~\eqref{eq:Nep}.
%%%%%%%%%%%%%%%%%%%%%%%%%%%%%%%%%%%%%%%%%%%%%%%%%%%%%%%%%%%%%%%%%%%%
\section*{Acknowledgment}
The author acknowledges the support from the \textit{Mathematical Physics Laboratory} of the \textit{Centre de Recherches Math\'ematiques}, through a postdoctoral scholarship. This research was also funded by Consejo Nacional de Ciencia y Tecnolog\'ia (Mexico), grant number A1-S-24569. 
%%%%%%%%%%%%%%%%%%%%%%%%%%%%%%%%%%%%%%%%%%%%%%%%%%%%%%%%%%%%%%%%%%%%

%%%%%%%%%%%%%%%%%%%%%%%%%%%%%%%%%%%%%%%%%%%%%%%%%%%%%%%%%%%%%%%%%%%%

\end{document}